\documentclass[aip,graphicx,preprint]{revtex4-1}
\usepackage[USenglish]{babel} 
\usepackage[utf8x]{inputenc}
\usepackage{lmodern} 
\usepackage{graphicx}
\usepackage{amsmath}

\newcommand{\ii}{\text{i}}
\newcommand{\neff}{n_\text{eff}}

\newcommand{\SnV}{SnV$^-$ center{ }}

\newcommand{\SnVs}{SnV$^-$ centers{ }}
\newcommand{\PSat}{$P_\text{Sat}${ }}
\newcommand{\ISat}{$I_\text{Sat}${ }}
\newcommand{\gtwo}{g^{(2)}(0)}

\begin{document}

\title[A cavity-based optical antenna for color centers in diamond]{A cavity-based optical antenna for color centers in diamond}

\author{Philipp Fuchs}
\author{Thomas Jung}
\affiliation{
Universität des Saarlandes, Fachrichtung Physik, Campus E2.6, 66123 Saarbrücken, Germany}
\author{Michael Kieschnick}
\author{Jan Meijer}
\affiliation{Universität Leipzig, Angewandte Quantensysteme, Linnéstraße 5, 04103 Leipzig, Germany}
\author{Christoph Becher}
\affiliation{
Universität des Saarlandes, Fachrichtung Physik, Campus E2.6, 66123 Saarbrücken, Germany}
\email{christoph.becher@physik.uni-saarland.de}

\date{\today}

\begin{abstract}
An efficient atom-photon-interface is a key requirement for the integration of solid-state emitters such as color centers in diamond into quantum technology applications. Just like other solid state emitters, however, their emission into free space is severely limited due to the high refractive index of the bulk host crystal. In this work, we present a planar optical antenna based on two silver mirrors coated on a thin single crystal diamond membrane, forming a planar Fabry-Pérot cavity that improves the photon extraction from single tin vacancy (SnV) centers as well as their coupling to an excitation laser. Upon numerical optimization of the structure, we find theoretical enhancements in the collectible photon rate by a factor of 60 as compared to the bulk case. As a proof-of-principle demonstration, we fabricate single crystal diamond membranes with sub-µm thickness and create SnV centers by ion implantation. Employing off-resonant excitation, we show a 6-fold enhancement of the collectible photon rate, yielding up to half a million photons per second from a single SnV center. At the same time, we observe a significant reduction of the required excitation power in accordance with theory, demonstrating the functionality of the cavity as an optical antenna.
Due to its planar design, the antenna simultaneously provides similar enhancements for a large number of emitters inside the membrane. Furthermore, the monolithic structure provides high mechanical stability and straightforwardly enables operation under cryogenic conditions as required in most spin-photon interface implementations.
\end{abstract}

\maketitle

\section{Introduction} 
In the past years, color centers in diamond involving a substitutional atom from group IV of the periodic table and a nearby vacancy (group IV vacancy centers \cite{Thiering2018}) gained strong interest for applications in quantum technologies \cite{Awschalom2018}, among them the negatively-charged silicon (SiV$^-$) \cite{Hepp2014,Becker2017}, germanium (GeV$^-$) \cite{Iwasaki2015,Palyanov2015}, and more recently also tin\cite{Iwasaki2017,Tchernij2017} (SnV$^-$) and lead vacancy center (PbV$^-$) \cite{Tchernij2017,Trusheim2019}. Their emission into the zero phonon line (ZPL) exceeds the well-studied negatively-charged nitrogen-vacancy center (NV$^-$) by one order of magnitude at room temperature\cite{Stadt}. Moreover, group IV vacancy centers possess an optically accessible electron spin \cite{Rogers2014a,Siyushev2017,Sukachev2017,Iwasaki2017,Becker2018}, rendering them well-suited building blocks for spin-photon-interfaces \cite{Awschalom2018}.

However, the high refractive index of diamond at visible wavelengths ($n=2.414$ at $620$ nm \cite{e6cvd}) induces a severe limitation on the photon extraction efficiency. 
Total internal reflection, trapping around $98~\%$ of the color center emission inside of bulk diamond, is a well-known problem \cite{Mi2020} also in competing systems such as semi-conductor quantum dots, or in classical light sources such as light emitting diodes (LED) \cite{Zhmakin2011}, where the refractive index of the host materials is typically even higher.
Consequently, there have been many proposals to increase the collectible photon rate from quantum emitters out of high index materials, most of them trying to shape the dielectric environment to circumvent total internal reflection. For color centers in diamond, non-resonant structures such as nanowires \cite{Babinec2010,Marseglia2018,Fuchs2018} or solid-immersion lenses \cite{Hadden2010,Marseglia2011,Riedel2014a,Yang2020} have been shown to be a feasible solution, yet they enhance the collectible photon rate only by one order of magnitude to around $30~\%$ of the total emission rate, and for only a small fraction of emitters with matching position and orientation. 
More sophisticated approaches involve resonant cavities, channeling the emission into one spatial and spectral mode and additionally enhancing the emission rate beyond the bulk emission rate via the Purcell effect\cite{Purcell}. This includes fiber-based Fabry-Pérot type cavities \cite{Burek2017,Benedikter2017,Dolan2018,Tomm2020,Ruf2020} as well as a plethora of integrated micro- and nanophotonic approaches \cite{Janitz2020,Mi2020}. Generally speaking, interfaces between solid state emitters and freely-propagating photons may be summarized as optical antennas \cite{Bharadwaj2009}: An optical antenna is defined as a device which is not only enhancing the collectible photon rate from an emitter, but \textit{vice versa} also enhances the coupling of external light to the emitter.
Whereas optical antennas have been realized with a multitude of designs\cite{Novotny2011}, the simplest structure may consist of a planar stack of dielectric layers as proposed by \textit{Lee et al.} \cite{Lee2011}. This approach has been demonstrated to be a promising option to enhance photon extraction for many emitters in the same structure, reaching outstanding collection efficiencies well above $95~\%$ for single molecules \cite{Lee2011}. The need for oil immersion microscopy to avoid total internal reflection, however, limits its applications to room temperature.
Circumventing this limitation has yet been shown to be possible: For color centers in a diamond membrane, a purely dielectric design for the extraction of the emitted photons can be achieved by placing a material with a higher refractive index in contact with the diamond \cite{Riedel2014a}. The diamond membrane thereby forms a leaky slab waveguide, directing the emission towards the higher index material, comparable to the design of \textit{Lee et al.}. The need for oil immersion can further be avoided by tailoring the higher index material as a macroscopic solid immersion lens \cite{Riedel2014a}. For ideal functionality, however, the color centers in the diamond membrane need to be placed centered with respect to the solid immersion lens. 
A similar design has also been demonstrated for quantum dots \cite{Ma2014}.
\par
In this work, we overcome the limitations of the designs just mentioned and return to an entirely planar design consisting of a dielectric slab with metal coatings. A comparable approach has recently been implemented by \textit{Checcucci et al.} \cite{Checcucci2017}, estimating collection efficiencies well above $50~\%$ for the emission of single molecules embedded in a crystalline matrix. It is based on two gold mirrors which are directly coated onto the host layer, yielding an intrinsic mechanical stability. The thickness of the host layer as well as of the mirrors is chosen to enhance the beaming of the molecule's emission into a narrow lobe in the far field. Just recently, an adaption to quantum dots has been reported by \textit{Huang et al.} \cite{Huang2021}, yielding a collection efficiency of around $19~\%$ with an air objective of NA = 0.85.
Although there exist theoretical \cite{Galal2017,Soltani2019} as well as first experimental studies \cite{Galal2019} to adapt this design also to color centers in diamond, a thorough demonstration is by now missing. 
\\
Thus, we start in Section \ref{sec:theoretical_description} with a detailed model of a dipole emitting in a planar metallo-dielectric stack based on a diamond membrane, identifying the design most generally as a monolithic Fabry-Pérot cavity.
In contrast to the studies mentioned above, we optimize the design towards a high absolute photon rate from a single \SnV in the diamond membrane, collectible with an air objective. We show that for certain thicknesses of the involved layers, the cavity enhances the collectible photon rate from single emitters inside the diamond membrane as well as the coupling of an excitation laser to these emitters, satisfying the definition of an optical antenna.
The theoretical description is followed by an optimization of the free parameters, revealing the main fabrication challenge: Whereas single molecules and quantum dots can be deposited in a bottom-up-approach, thin diamond membranes are typically fabricated in a top-down-approach involving plasma etching of high quality bulk diamond \cite{Challier2018,Heupel2020} and subsequent doping with color centers. Thus, we continue with briefly describing the fabrication of the diamond membrane as well as of the metallo-dielectric layers (Sec. \ref{sec:fab}). 
To verify the functionality of the fabricated antenna, we perform a spectroscopic investigation of single \SnVs prior to and after applying the coatings in Section \ref{sec:res}. Yielding a nearly 6-fold enhancement of the single photon emission rates in accordance with theory, together with a severe reduction of the required excitation power, we demonstrate the operation of the fabricated cavity in terms of an optical antenna as defined above.
Especially for the implementation of single photon sources or spin-photon interfaces, a monolithic and thus inherently stable optical antenna together with an intrinsically high scalability and low technical overhead paves the way towards the usage of color centers in diamond as versatile building blocks for present and future quantum technologies. 
\section{Theoretical description}
\label{sec:theoretical_description}
\begin{figure*}[h]
\centering
\includegraphics[width=\linewidth]{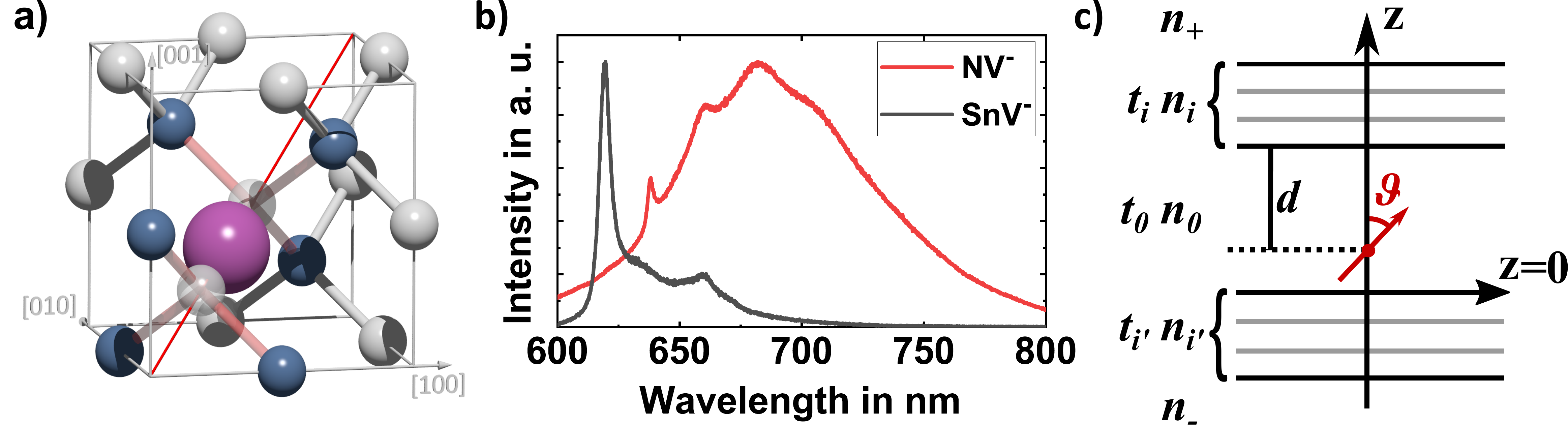}
\caption{(a) Upon formation of an SnV$^-$ center, the tin atom (purple) moves to an interstitial position between two lattice sites, forming a split-vacancy configuration. The carbon atoms contributing to the color center are colored blue. 
(b) The photoluminescence spectrum of the \SnV in bulk diamond shows a prominent zero-phonon line at around 620 nm, in contrast to the dominant phononic side band of the NV$^-$ center emission shown in red. (c)
For the nanophotonic calculations, we model the \SnV as an electric point dipole and place it at depth $d$ inside a diamond host layer of thickness $t_0$ and refractive index $n_0$. The dipole spans an angle $\vartheta$ with the z-axis and its emission is described by a plane wave expansion approach. The host layer is situated between two layer systems with refractive indices $n_i$ and thicknesses $t_i$ for the layers above and $n_{i'}$ and $t_{i'}$ for the layers below, respectively. The whole stack is embedded within semi-infinite layers on both sides with refractive index $n_\text{+}$ and $n_\text{-}$, respectively. We assume the collection optics to be placed in the upper half space with refractive index $n_\text{+}$. }
\label{fig:Basics}
\end{figure*}

In the following, we focus entirely on the \SnV with its comparably high photon rate already out of bulk diamond, typically reaching on the order of $10^4 - 10^5$ counts per second (cps) at the detectors (compare saturation statistics in Figure \ref{fig:finalStatistics}).
The atomic structure of the \SnV is depicted in Figure \ref{fig:Basics} (a). Made up of a substitutional tin atom and an adjacent vacancy, the tin atom moves to an interstitial position between the two lattice positions, forming a split-vacancy configuration \cite{Thiering2018,Bradac2019}. In contrast to the NV center, this gives rise to an inversion symmetry, yielding a strongly reduced sensitivity of the electronic levels to external perturbations, in particular to electric fields.
The photoluminescence (PL) spectrum of the \SnV consists of the ZPL at 620 nm, followed by a phononic side band (PSB) extending to over 700 nm. In previous work, we have thoroughly studied the temperature-depended Debey-Waller factor, quantifying the branching ratio between ZPL and PSB as 30:70 at room temperature \cite{Stadt}. This exceeds the corresponding value of the NV center by one order of magnitude, rendering the \SnV a promising candidate for the implementation of efficient room temperature single photon sources.
An exemplary spectrum of an \SnV in bulk diamond at room temperature is shown in Figure \ref{fig:Basics} (b).
\par
In the next section, we want to discuss the emission characteristics of a single \SnV situated inside a planar metallo-dielectric stack.
From a photonic point of view, the \SnV can be modelled as an electric point dipole with a dominant dipole axis along one of the high symmetry $\langle 111 \rangle$ axes of the diamond lattice \cite{Hepp2014,Stadt}, depicted as red solid line in Figure \ref{fig:Basics} (a).
An electric dipole radiating inside a planar metallo-dielectric stack is sketched in Figure \ref{fig:Basics} (c): We assume the dipole to be embedded in a diamond layer of thickness $t_0$ and refractive index $n_0$, enclosed by two arbitrary layer systems. 
The whole stack is embedded within semi-infinite layers on both sides. We further assume the collection optics to be placed in the half space above the dipole with real refractive index $n_\text{+}$, allowing us to collect all emission from the dipole leaving the stack in positive z-direction up to a detection angle $\theta_\text{NA}$ defined by the NA of the collection optics via $\text{NA} = n_\text{+} \cdot \sin(\theta_\text{NA})$. We call this half space the collection half space.
\par 
The dipole emission in such a stack can be treated completely by classical electrodynamics, which has been shown multiple times in literature \cite{Lukosz1979,Polerecky2000,Arnoldus2004,Baets2008}:
Starting with a plane wave expansion of the emitted electric field, the corresponding plane waves and evanescent fields are propagated through the layers using generalized Fresnel coefficients for the layer system above and below the diamond layer obtained via a transfer matrix method \cite{Yeh1988}. In the semi-infinite half spaces, we perform a far field transformation using the method of stationary phase \cite{Mandel1995}. With this formalism, we can derive expressions for the radiated far fields that can be evaluated numerically.
\subsection{Optical antenna working principle}
\begin{figure*}[h]
\centering
\includegraphics[width=\linewidth]{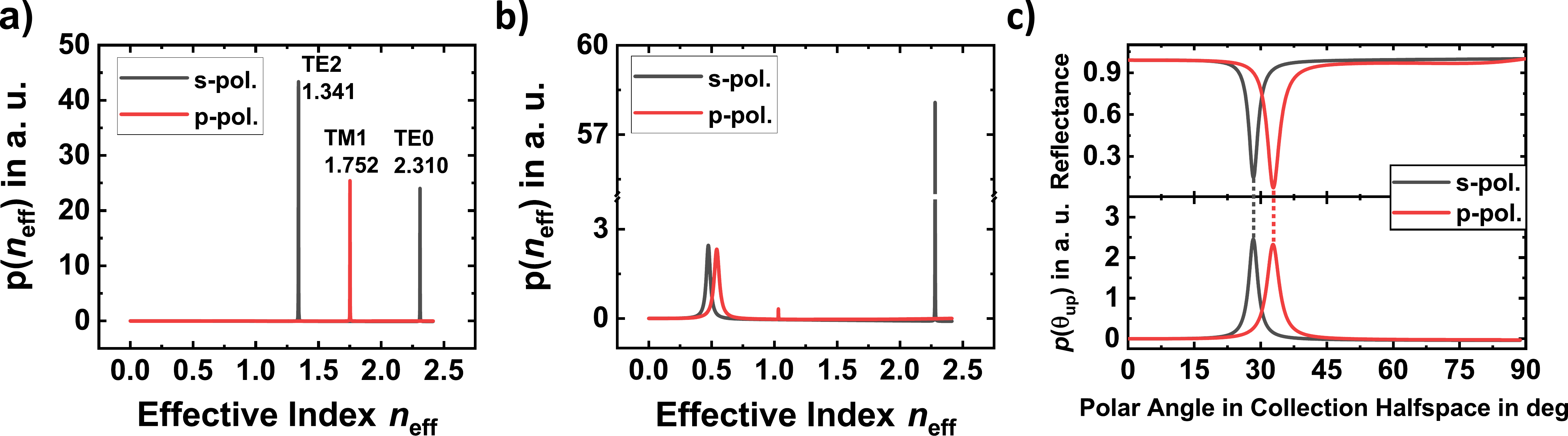}
\caption{(a): The angular power emission spectrum $p(\neff)$ of a dipole placed in the middle of a diamond slab with $t_0 = 350$ nm surrounded by air. The peaks correspond to a coupling of the dipole to guided modes. (b) Going from dielectric total internal reflection to metallic mirror (silver) reflection introduces higher losses, but also the possibility for modes with $\neff < 1$ to leak from the diamond slab towards the upper half space, assuming the upper silver layer to be thin enough. (c) Using Snell's law, we can calculate the polar angles $\theta_\text{up}$ at which these modes propagate as plane waves in the collection half space. By evaluating Fresnel's equations for plane waves incident on the stack from the collection half space, we can additionally probe the reflectance of the stack.}
\label{fig:WorkingPrinciple}
\end{figure*}

The optical antenna design is based on a thin ($t_0 < 1~\mu$m) diamond membrane with refractive index $n_0 = 2.414$ at 620 nm \cite{e6cvd}, which itself can be modelled as a slab waveguide when surrounded by air ($n_\text{+} = n_\text{-} = 1.0$). Consequently, most of the emission from color centers is trapped inside the membrane due to total internal reflection at the diamond-air interfaces. The corresponding guided modes can be calculated by solving the transcendental eigenvalue equations\cite{Bures2009}.
Another possibility to reveal the guided modes is to virtually place an electric point dipole inside the slab and look at its emitted power, which is dependent upon the local density of states (LDOS): In a homogeneous medium with refractive index $n_0$, the emitted power is given as $P_\text{hom} = n_0 \cdot P_0$ with $P_0$ the emitted power in vacuum. An arbitrarily shaped inhomogeneous environment such as a slab waveguide gives rise to a modified LDOS and thus a change in the emitted power\cite{PrinciplesOfNanoOptics}. This change can be modeled by splitting the total power $P_\text{tot}$ emitted by the dipole in a sum of the homogeneous power $P_\text{hom}$ and an additional contribution $P_\text{inhom}$. Furthermore, $P_\text{inhom}$ can be expressed in terms of a plane wave expansion as shown in equation \eqref{eq:planewave}.
\begin{align}
P_\text{tot} &= P_\text{hom} + P_\text{inhom} \nonumber \\ 
&= P_\text{hom} + \int_{0}^{\infty} p(k_\parallel) \text{d}k_\parallel \nonumber \\
&= P_\text{hom} + \underbrace{\int_{0}^{k_0 n_0} p(k_\parallel) \text{d}k_\parallel}_\text{plane waves} + \underbrace{\int_{k_0 n_0}^{\infty} p(k_\parallel) \text{d}k_\parallel}_\text{evanescent fields}
\label{eq:planewave}
\end{align}
The integrals in equation \eqref{eq:planewave} break down the electric field of the dipole into a series of plane waves and evanescent fields. Each component of the series is propagating in a different direction which is determined by the parallel component $k_\parallel = k n_0 \sin(\theta)$ of the wave vector with $k = 2\pi / \lambda$ and $\lambda$ the vacuum wave number and wavelength, respectively. 
Thus, $k_\parallel$ is directly related to the polar emission angle $\theta$ and we call the function $p(k_\parallel)$ the angular power emission spectrum.
Defining $\neff = n_0 \sin(\theta)$, we find the figure $\neff$, which is well-known from guided wave optics and is called effective index of the guided modes.
As $k_\parallel$ is a rather unhandy value, we evaluate $p(\neff)$ instead of $p(k_\parallel)$, enabling us to directly extract the effective index of the guided modes.
As an example, Figure \ref{fig:WorkingPrinciple} (a) shows $p(\neff)$ for a diamond slab ($t_0 = 350$ nm, $n_0$ = 2.414 at $\lambda = 620$ nm\cite{e6cvd}) surrounded by air and a dipole placed exactly in the middle of the slab ($d=175$ nm), oriented parallel to the interfaces ($\vartheta = 90^\circ$).
At certain $\neff$, the LDOS and thus the emission of the dipole is enhanced due to the presence of a guided mode, yielding a peak in $p(\neff)$. For a loss-less waveguide, these peaks correspond to poles of $p(\neff)$. To avoid this numerically, we introduce a small absorption of $\kappa = 5\cdot 10^{-4}$ to the refractive index of the diamond layer.
\par
If the layer above and below the diamond slab is chosen to be air with refractive index $n_\text{+}=n_\text{-} = 1$, the range $1 < \neff < n_0$ corresponds to plane waves of the expansion which are trapped in the slab. In this regime, we see peaks corresponding to the guided modes, delivering the same effective indices as the direct approach of solving the transcendent equations for a slab waveguide. By evaluating $p(\neff)$, however, we can extract to which modes the dipole actually couples, as due to the chosen dipole orientation and position, it does not couple to all existing guided modes.
\par
These considerations also hold, when we introduce silver coatings ($n_1 = n_{1'} = 0.05 + 4.21\text{i}$ at 620 nm\cite{McPeak2015}) on both sides of the diamond slab. Instead of having total internal reflection, the diamond-silver interfaces possess a high reflectance independently of the angle of incidence, leading again to modes in the diamond slab which can also be calculated via suitable eigenvalue equations \cite{Chen2000}. In the absence of a critical angle, all peaks with $\neff < n_0$ in $p(\neff)$ correspond to guided modes of the stack.
Reducing the thickness of the top silver layer to values well below 100 nm, the layer becomes transparent enough to let the formerly guided modes leak out of the diamond slab trough the thin silver layer into the collection half space with $n_\text{+}=1.0$. This effectively implies that modes with $n_\text{eff} < n_\text{+}=1.0$ will be converted to free-propagating radiation at well-defined polar angles in the collection half space beyond the thin silver layer. Consequently, these modes are often named leaky modes.
The resulting angular power emission spectrum for an upper silver mirror with $t_1 = 50$ nm thickness is shown in Figure \ref{fig:WorkingPrinciple} (b). The lower silver mirror is kept thick and thus opaque with $t_{1'} = 300$ nm.
The two peaks in $p(\neff)$ at $\neff=0.475$ (s-pol.) and $\neff = 0.542$ (p-pol.) correspond to leaky modes. Using Snell's law, we can calculate the polar angles $\theta_\text{up}$ at which these modes propagate as plane waves in the collection half space. Additionally, we can probe the reflectance of the stack by evaluating Fresnel's equations for plane waves incident on the stack from the collection half space. This is shown in Figure \ref{fig:WorkingPrinciple} (c), where we see sharp drops in reflectance at angles of incidences coinciding with the propagation angles of the plane waves corresponding to the leaky modes.
Consequently, we can identify the stack with a plane-parallel Fabry-Pérot cavity, because we can probe it independently of the dipole emission via reflectance calculations. For a matching cavity length, the cavity alters the dielectric environment and thereby the LDOS and enables a dipole inside it to couple to the leaky modes, which are efficiently transferred to free propagating radiation in the collection half space. Vice versa, light sent onto the cavity under matching angles is efficiently coupled to the cavity and thus to the dipole. This matches well the definition of an optical antenna\cite{Bharadwaj2009}. In the next section, we will determine the ideal cavity length for an efficient antenna operation as part of a general optimization of all free parameters.
\par
The planar design considered here has already been discussed as a cavity in literature and is thus conceptually well understood, especially in the context of enhancing light extraction from LEDs \cite{Benisty1998,Baets2003,Baets2008}. 
We, however, want to point out that some of the underlying theoretical concepts are much more sophisticated than the presented model. As the transmission through the thin silver layer is the major loss source of the cavity, the leaky modes can no longer be taken as confined between the mirrors, which is a major assumption in resonator theory. Instead, they span trough the semi-transparent mirror towards infinity, defining the system as a so-called open cavity. Such open cavities have already been discussed on a fundamental level \cite{Dutra1996}, identifying Fox-Li quasimodes as the exact solutions.
However, as will be shown in the next sections, the model presented above describes the experimental results successfully without the need for specific assumptions on the cavity modes.
\subsection{Optimization}
\label{sec:optimization}
After discussing the basic concept of the design, it becomes obvious that the thickness of the diamond as well as of the thin silver layer can be optimized to enhance the coupling of an \SnV to the cavity, enabling the functionality as an optical antenna.
To quantify the brightness of a single photon emitter driven by an off-resonant continuous wave laser, a well-suited figure of merit is given by the collection factor $\xi$ in equation \eqref{eq:SingleCollFac}:
\begin{equation}
\xi = \frac{\Gamma_\text{NA}}{\Gamma_\text{hom}}
\label{eq:SingleCollFac}
\end{equation}
$\Gamma_\text{NA}$ is the far field photon rate into a solid angle defined by the NA of the collection optics, whereas $\Gamma_\text{hom}$ is the radiative decay rate of an electric dipole transition inside a homogeneous medium. Thus, we have $\Gamma_\text{hom} = n_0 \Gamma_0$ with $\Gamma_0$ the vacuum emission rate. 
The collection factor $\xi$ hence defines a handy and comparable value for the absolute collectible photon rate $\Gamma_\text{NA}$ under continuous wave excitation. 
Due to an enhanced LDOS, we may also find $\xi>1$, which can be attributed to a lifetime reduction of the emitter (Purcell effect\cite{Purcell}).
\par
As we calculate classical electromagnetic fields and optical powers, but aim at investigating quantum mechanical emission rates of single quantum emitters, we use the following relation:
\begin{equation}
\frac{\Gamma_\text{tot}}{\Gamma_\text{hom}} = \frac{P_\text{tot}}{P_\text{hom}}
\label{eq:classicQED}
\end{equation}
Equation \eqref{eq:classicQED} links both classical optical powers $P$ and quantum-mechanical rates $\Gamma$, allowing us to restrict the investigations to purely classical calculations.
It can be derived by comparing the classical Green's function of an electric point dipole and the quantum mechanical decay rate of an electric dipole transition \cite{PrinciplesOfNanoOptics}. 
\par
Next, we want to probe the theoretical limits of the design in terms of the just defined figure of merit $\xi$.
The open parameters are the thickness of the diamond membrane $t_0$, the depth of the dipole below the thin silver layer $d$, and the thickness of the thin silver layer $t_1$ on top of the diamond membrane. As we place the collection optics in the half space above the thin silver layer, we leave the thickness of the lower silver layer fixed at $t_{1'} = 300$ nm to avoid any radiation from leaking to the lower half space. The collection optics is assumed to be an air objective with a collection angle of around $53^\circ$ defined by the $\text{NA}=0.8$.
Additionally, we introduce a silica layer of thickness $t_2$ on top of the thin silver layer to prevent the latter from corroding over time. 
\par
We carry out the calculations with a self-developed implementation of the plane wave expansion using \textit{python}. For the optimization we employ a particle swarm optimization algorithm (\textit{pyswarm}) in combination with a classical optimization algorithm (\textit{scipy}, \textit{L-BFGS-B}).
For the refractive indices, we use literature values for $\lambda = 620$ nm, yielding $n_1=n_{1'}=0.05+4.21\ii$ for silver \cite{McPeak2015}, $n_2=1.464$ for silica \cite{Rodriguez-deMarcos2016} and $n_0=2.414$ for diamond  \cite{e6cvd}. 
A dipole orientation parallel to the interfaces ($\vartheta = 90^\circ$) is the most efficient configuration by symmetry argumentation. We name this optimal case in the following case (I). In addition, we also perform the optimization with a dipole polar angle of $\vartheta=54.7^\circ$, corresponding to the actual orientation of the \SnV symmetry axis and thus the dipole axis in commonly-available (001)-oriented diamond (case (II)). The results of both optimizations are summarized in Table \ref{tab:Optimization}.
\par
\begin{table}[h]
\centering
\caption{Optimization results for different cases as specified in the main text. Values for $t_0$, $t_1$, $t_2$ and $d$ are given in nanometer.}
 \begin{tabular}{|c|c|c|c|c|c||c|}
 \hline
   & $\vartheta$ & $t_0$ & $d$ & $t_1$  & $t_2$ & $\xi$ \\
  \hline
  (I) & $90.0 ^\circ$ & $86.5$ & $42.9$ & $42.4$ & $107.6$ & $2.01$ \\
  (II) &$54.7 ^\circ$ & $86.5$ & $42.9$ & $42.4$ & $107.6$ & $1.34$\\
  \hline
  (III) &$54.7 ^\circ$ & $609.2$ & $27.5$ & $24.9$ & $107.7$ & $0.28$\\
  \hline
\end{tabular}
\label{tab:Optimization}
\end{table}
\par
The values in case (I) lead to the highest $\xi$ that is theoretically achievable, given the optical properties from literature. With $\xi =  2.01$, we get a 87-fold enhancement of the collectible photon rate compared to the bulk case with $\xi = 0.023$. For the realistic case (II) of a non-parallel dipole orientation, $\xi = 1.34$ still corresponds to a 58-fold increase.
Having identified the design as a cavity in the previous section, it is intuitive that the optimization tends to a very thin diamond membrane. More precisely, the stack forms a $\lambda/2$ cavity in case (I) and (II), considering the effective index of the leaky modes and the correct penetration depth $d_\text{pen}$ into the mirrors \cite{Ma2007} according to equation \eqref{eq:resonanceCondition}.
\begin{equation}
q \cdot \lambda / 2 = t_0 \cdot \sqrt{n_0^2 - \neff^2} + d_\text{pen} ~~~~ q = 1,2,3\dots
\label{eq:resonanceCondition}
\end{equation}
As an example, for case (I) we find $\neff = 0.33$ for the s-polarized mode and $d_\text{pen} = 52.1$ nm ($50.8$ nm) for its penetration depth into the upper (lower) stack, summing up to $309.8~\text{nm}$. This matches $\lambda/2 = 310~\text{nm}$ very well.
The thickness $t_2$ of the silica layer tends to $\lambda/{(4 \cdot n_2)}$, working as an anti-reflective coating.
\par
\begin{figure*}[h]
\centering
\includegraphics[width=\linewidth]{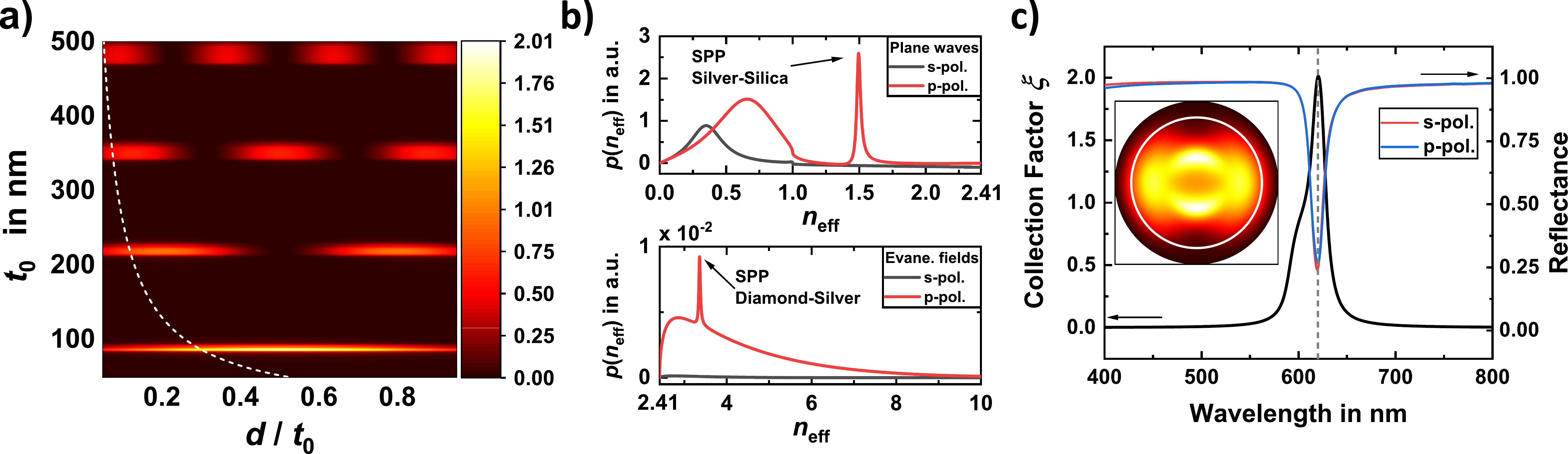}
\caption{All plots shown are based on the parameters of case (I) in table \ref{tab:Optimization}. (a) Sweeping the thickness $t_0$ of the diamond and the depth $d$ of the dipole inside, we find the collection factor $\xi$ to peak whenever we hit a thickness for which the cavity becomes resonant to the dipole emission wavelength according to equation \eqref{eq:resonanceCondition}. Additionally, the dipole must be situated in an electric field node of the corresponding cavity mode. The white dashed line describes the actual depth of the fabricated color centers limited by the implantation energy. (b) The angular power emission spectrum $p(\neff)$ reveals the leaky modes at $\neff < 1$ as well as the main loss channels that can be attributed to surface plasmon polaritons (SPP). (c) The spectral width of the cavity resonance dictates possible excitation wavelengths that must lie within. The otherwise high reflectance prevents an efficient coupling of the excitation light. The reflectance is calculated for incidence angles in the same range as the emission angles that can be seen from the inset, showing the far field in the collection half space. The white ring indicates the collectible fraction given by the NA of the objective we employ.}
\label{fig:OptimizationResult}
\end{figure*}
To gain a deeper insight, Figure \ref{fig:OptimizationResult} (a) provides $\xi$ in dependence of $t_0$ and $h$ for case (I). As expected, high values of $\xi$, corresponding to the dipole being in resonance with the cavity, occur only for $t_0$ fulfilling equation \eqref{eq:resonanceCondition}, and the number of field nodes increases with $t_0$. 
We want to emphasize that in Figure \ref{fig:OptimizationResult} (a), we do neither plot the cavity resonances directly in terms of Purcell factor, nor do we show an actual field distribution. Shown is $\xi$ as a measure of the collectible photon rate. It becomes obvious that an enhanced emission is given only for a good coupling to the cavity, thus $\xi$ maps the field distribution inside the cavity.
For thicker diamond layers, additional modes with $\neff > 1.0$ start to appear. As mentioned above, modes with $\neff > 1.0$ are confined in the slab and do not contribute to $\xi$. Additionally, the coupling to the leaky modes reduces because of a reduced field enhancement for thicker cavities, comparable to an increasing mode volume, leading to a reduction of $\xi$.
\par
Figure \ref{fig:OptimizationResult} (b) shows $p(\neff)$ for case (I), indicating its physical limitations: The metallic mirrors introduce absorption losses as well as coupling to plasmonic resonances, i.e. surface plasmon polaritons (SPP). SPPs can occur either at the silver-silica interface or at one of the diamond-silver interfaces. The former are excited by plane waves, compare the peak at $\neff = 1.5$ in Figure \ref{fig:OptimizationResult} (b). The latter can only be excited by the near field of the dipole and are thus visible in the evanescent part of the angular power emission spectrum.
Circumventing these plasmonic resonances is possible by increasing the optical distance of the dipole to the silver layers, yet the color center creation method limits us in this work: We utilize ion implantation, yielding a mean implantation depth for the tin ions of only $d=27.5$ nm as we will discuss in the next section.
In future implementations, this limitation can be overcome by either implanting at higher energies or by introducing buffer layers with low refractive index such as silica between the diamond and the silver layers \cite{Galal2017,Huang2021}. This, however, requires a more precise control over the layer deposition and, more challenging, a thinner diamond membrane.
\par
A more severe limitation becomes evident when looking at the width of the cavity resonance for case (I) shown in Figure \ref{fig:OptimizationResult} (c). As designed, $\xi$ peaks around the ZPL at 620 nm. Although the resonance is comparably broad ($23$ nm \textit{FWHM}), it implies that we have to choose an excitation wavelength inside it, because the cavity will reflect most of the incident light outside. A common experimental situation for off-resonant excitation of many color centers in diamond is using a green laser; in this work we employ a laser emitting at $516$ nm. Consequently, for such a situation we have to find a cavity length at which both the green excitation light as well as the ZPL of the SnV$^-$ center are resonant.
Sweeping $t_0$ while keeping the other parameters fixed as in case (I), we find $t_0 = 609$ nm as the smallest thickness at which the cavity is resonant for both wavelengths, defining this thickness as the working point of the cavity as an optical antenna for the off-resonant excitation. 
We redo the optimization to check whether for this thicker membrane and the limited implantation depth, other thicknesses of the thin silver and silica layer yield a better enhancement. Indeed, as can be seen from case (III) in table \ref{tab:Optimization}, the optimal silver layer thickness decreases here to $t_1 = 24.9$ nm, whereas the silica thickness stays nearly the same. With $\xi=0.28$, the enhancement is still 12-fold, yet we loose a factor of 5 compared to case (II) with resonant excitation. To demonstrate the functionality, the off-resonant excitation should however provide a measurable change in the collectible photon rate. In future experiments, we may explore resonant excitation to benefit from the full potential of the design.
Lastly, it is worth mentioning that in all three cases, the cavity is only resonant for excitation light incident under the correct angle. This can be seen from the reflectance curves in Figure \ref{fig:OptimizationResult} (c), which show a dip in reflectance overlapping with the peak in $\xi$.
Here, the reflectance is calculated assuming an angle of incidence covering the broad emission angle distribution that is inferable from the far field plot in the inset. The white ring in the far field plot indicates the collection angle covered with our experimentally-used NA = 0.8 air objective. As the white ring includes most of the emitted light and the excitation is performed via the same objective, we can vice versa be sure to cover the required angles of incidence in excitation.
\section{Experimental Methods}
\label{sec:fab}
\subsection{Fabrication}
As starting material, we use commercially-available single crystal diamond, grown by chemical vapor deposition (\textit{Element Six}, \textit{electronic grade}). These bulk $(001)$-oriented diamond plates are subsequently polished (\textit{Delaware Diamond Knives}) to a final thickness of 30 µm and cut to lateral dimensions of $2~\text{mm}\times 4~\text{mm}$.
After cut and polish, we perform a solvent clean (2x5 min acetone + 2x5 min isopropyl alcohol with ultrasonic support) to remove residual contaminants from polishing. 
The solvent clean is followed by an annealing step (4 h at 1200 $^\circ$C and $<10^{-6}$ mbar) to reduce potential damage in the diamond. 
Prior to annealing, we purge the furnace with pure nitrogen. The annealing starts with a temperature ramp over 36 h to maintain pressures well below $10^{-6}$ mbar at any time.
During annealing, the beforehand colorless and transparent diamond turns grayish because of non-diamond carbon forming at the surface.
To remove this non-diamond carbon, we perform a tri-acid cleaning (1 part nitric acid : 1 part sulfuric acid : 1 part perchloric acid, boiling at 500 $^\circ$C) and a subsequent oxidation step (2 h a 450 $^\circ$C in air at ambient pressure), ending up again with a transparent diamond. 
After this initial processing, the diamond plates are ready for the first dry etching step, which is performed in a reactive ion etching machine (RIE, \textit{Roth \& Rau MicroSys 350}) according to the recipe described by \textit{Jung et al.} \cite{Jung2016}. 
Prior to etching, we glue the diamond plates to a clean silicon substrate and cover them with a quartz mask. Windows in the masks define the places where the diamond plates will be thinned down by the RIE process. To avoid local overetching and thus the formation of trenches, the windows in the masks possess angled sidewalls as described by \textit{Challier et al.} \cite{Challier2018}. 
We remove around 19 µm in the non-masked regions of the diamond plates. After the etch, the diamond plates possess well-defined windows and the mask can be removed. The etched diamond plates are subsequently taken to an ion accelerator to perform the implantation of tin ions.
\par
Limited by the analyzing magnet of the accelerator, we implant the tin ions at 80 keV, yielding a depth of 27.5 nm with an axial straggle of 5 nm according to the Monte-Carlo simulation \textit{stopping ranges of ions in matter} (\textit{SRIM} \cite{Ziegler1985}). For the simulation, we assumed the density of diamond to $3.52~\text{g}\,\text{cm}^{-3}$. We implant at a fluence of $10^9$ Ions/cm$^2$, equal to $10$ Ions/$\mu$m$^2$.
After the implantation, we remove the diamond plates from the silicon substrate and repeat the cleaning and annealing steps as outlined above to finally generate the SnV$^-$ centers. 
\par
To yield diamond membranes with thicknesses below 1 $\mu$m in the windows, we perform a series of short etching steps from the not yet etched backside of the diamond plates without the need for masks. These etching steps are done with an inductively-coupled plasma (ICP, \textit{Oxford Plasmalab 600}) supported RIE process using an oxygen plasma \cite{Challier2018}. 
Every etching step is followed by a white light microspectrometry measurement (\textit{A.S. \& Co. PDA Vis}) to determine the thickness of the membranes. 
\par
Lastly, we apply the coatings via electron beam evaporation (\textit{Pfeiffer Classic 500 L}). For silver as well as silica evaporation, we use a deposition rate of 1 \AA$/$s. Together with the diamond, we always add silicon substrates as control samples to the evaporation chamber to check the resulting thickness via a lift-off process and a subsequent measurement via atomic force microscopy (\textit{Park AFM XE-70}). Additionally, optical properties of the evaporated layers are determined via spectroscopic ellipsometry (\textit{HORIBA Jobin Yvon UVISEL-NIR}) of these control samples.
This thorough characterization enables us to feed the model with precisely-measured thicknesses as well as dispersion and absorption data of the actually fabricated device. Only for the diamond itself, we take the dispersion values provided by \textit{Element Six}\cite{e6cvd}.
\subsection{Spectroscopic characterization}
To characterize the diamond membrane and the \SnVs inside prior to and after coating, we perform spectroscopy using a home-build confocal laser-scanning PL microscope. For excitation, a continuous-wave diode laser emitting at 516 nm (\textit{Toptica iBeamSmart}) is used, which is coupled to a single mode fiber for spatial cleaning of the beam. For spectral cleaning, a narrow bandpass filter is placed behind the fiber outcoupler. The collimated laser beam is guided via a dichroic mirror (567 nm cut-on wavelength) to an air objective (\textit{Olympus LMPlan FL 100x}, NA = 0.8), which is also used to collect the PL from the sample. The sample itself is mounted under ambient conditions on linear tables. The PL is separated from the reflected laser light by the dichroic mirror and an additional 610 nm longpass filter suppresses the laser light. A 650 nm shortpass filter may be added to further narrow the detection bandwidth.
The filtered PL is coupled into a single mode fiber, working as confocal pinhole, and can be sent to a Hanbury-Brown and Twiss setup (HBT) for photon statistics and generation of PL maps, or to a grating spectrometer. The HBT setup involves a 50:50 beamsplitter and two avalanche photon diode detectors (\textit{PerkinElmer SPCM-AQR-13}), yielding a measured total detection time jitter of around 700 ps, which we take into account when fitting measured coincidence rates.
\section{Results}
\label{sec:res}
\subsection{Bare diamond membrane}
After several etching cycles, we end up with a diamond membrane possessing an average thickness well below 1 $\mu$m. In the white light microscopy image in Figure \ref{fig:BareMembraneLM} (a), thin-film interference can be observed. The orientation of the fringes indicates a thickness gradient in the membrane, which we estimate to around 1.8 nm change of thickness per µm change in lateral position in the direction of steepest slope. Although this gradient can be seen as a fabrication imperfection, it effectively provides us a varying cavity length.
\par
\begin{figure*}[h]
\centering
\includegraphics[width=\linewidth]{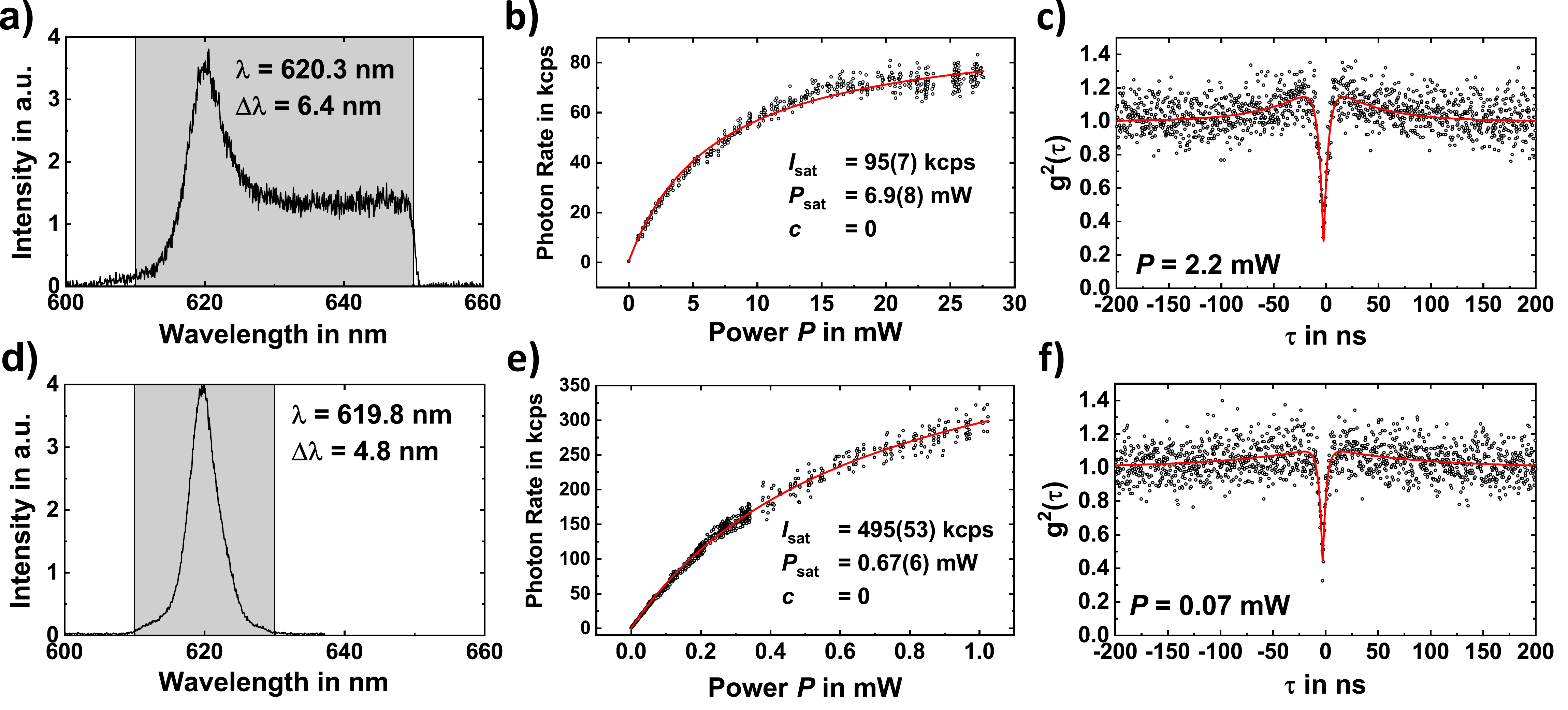}
\caption{An exemplary PL spectrum (a), saturation measurement (b) and photon autocorrelation measurement (c) of an \SnV in the bare diamond membrane. The PL spectrum reveals the characteristic ZPL at 620 nm with the adjoined PSB and a sharp drop at 650 nm because of the filter edge. The red lines are fits to the data. Although we cannot fit the linear background contribution to the saturation measurement, the residual terms describe the saturation well enough to provide comparable values for saturation count rate and power. The non-perfect dip in the autocorrelation measurement can be explained by uncorrelated background fluorescence and the detection time jitter. After applying the coatings, we find emitters close to the working point at $t_0 = 609$ nm that show comparable dips in photon autocorrelation (f) as for the uncoated membrane. The saturation measurement (e) reveals the functionality of the cavity at the working point as an optical antenna, as we need less excitation power to get a higher photon rate from the emitter. The spectrum (d) still shows the ZPL of the SnV$^-$ center, which at this position matches well the cavity resonance wavelength. The spectrum by accident was taken with a different detection window, while saturation and autocorrelation measurement have been performed with the default detection window from 610 - 650 nm.}
\label{fig:BareMembraneMeas}
\end{figure*}
Using microspectrometry, we can measure the average thickness in small areas of the membrane. This together with a model to calculate the color-depended reflectance of the membrane enables us to match the interference fringes with certain thicknesses of the membrane.
Prior to applying the coatings, we look for individual \SnVs in the thinnest part of the diamond membrane to measure their saturation properties. 
We find many bright spots, yet not all of them show a clear signature of single photon emission, i.e. a measured $\gtwo$ below $0.5$, where the deviation to a perfect dip is fully explained by detection time jitter and uncorrelated background emission from the diamond surface. Emitters with $ 0.5 \leq \gtwo < 1.0$ are not taken into account. 
Figures \ref{fig:BareMembraneMeas} a), b) and c) show a set of measurements performed on the same single \SnV in the bare membrane, representative for the emitters we include in the statistics.
In the spectrum (a), a clear signature of \SnV emission is given by the ZPL at 620 nm, which has a linewidth of around $6$ nm at room temperature. 
Recording the detected photon rate $I(P)$ for increasing excitation power $P$, we find a saturation behavior as shown in Figure \ref{fig:BareMembraneMeas} (b). We fit $I(P)$ with a sum of three contributions\cite{Neu2012a}: A constant contribution $D$ accounting for the detector dark counts (500 cps in total), a linear contribution $c\cdot P$ accounting for uncorrelated background fluorescence, and a non-linear saturation term modeling the actual PL of the \SnV yielding equation \eqref{eq:SatMeas}.
\begin{equation}
I(P) = \frac{I_\text{sat}\cdot P}{P + P_\text{sat}} + c \cdot P + D
\label{eq:SatMeas}
\end{equation}
For most of the \SnVs we measure, however, we are not able to satisfactorily fit the data with a positive linear background contribution constant $c$ although we observe background fluorescence. The fits tend to the nonphysical regime of $c<0$, thus we set $c=0$ fixed in all fits. We assume that this behavior originates from a complex charge state dynamic at high excitation powers, yet the underlying mechanism is still unclear. 
The values for \PSat and \ISat are however reliable and suitable for relative comparison of emitters in the bare membrane and in the antenna, as can be seen from the fit (red line) in Figure \ref{fig:BareMembraneMeas} (b) and (e), modeling the measured data in both cases well.
For moderate excitation powers, we measure a photon autocorrelation with the HBT setup as shown in Figure \ref{fig:BareMembraneMeas} (c), yielding in this example $\gtwo = 0.3$. The detection time jitter is taken into account in form of a convoluted fitting function.
Because of the missing possibility to estimate the background contribution directly from the saturation measurement, we independently estimate the background by integrating a PL spectrum at a position nearby the emitter. For the exemplarily-shown  $g^{(2)}(\tau)$ in Figure \ref{fig:BareMembraneMeas} (c), the residual coincidences from the fit yield a ratio of the emitter PL to the total PL (emitter and background) of 0.87(4), in perfect agreement to a value of 0.88 calculated from an integration of the corresponding spectra. 
\subsection{Membrane with applied coatings}
Besides the thickness of the diamond membrane, the final performance of the antenna strongly depends on the quality and thus the reflectance of the silver layers. 
Most probably due to an insufficient calibration of the evaporator, we do not reach the target thicknesses of the layers. We end up with $t_1 = 30$ nm of silver, followed by $t_2 = 128$ nm of silica for the upper layers. 
The thick silver layer on the back side has a thickness of $t_1' = 160$ nm.
\par
\begin{figure*}[h]
\centering
\includegraphics[width=.9\linewidth]{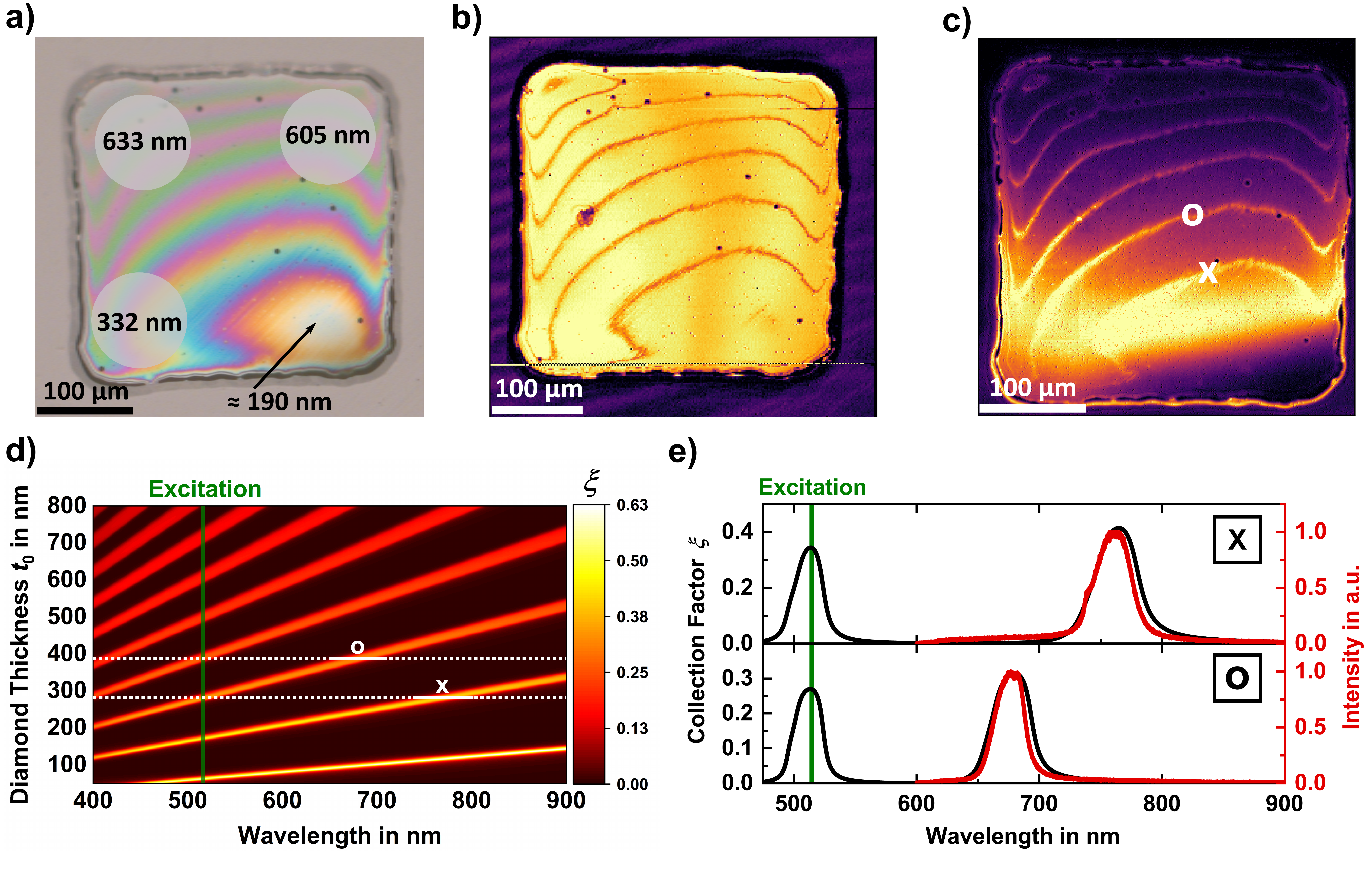}
\caption{(a) White light microscopy image of the final diamond membrane without coatings applied. The clearly visible thin film interference fringes indicate the gradient in the thickness of the membrane. Microspectrometry enables us to precisely determine the average thickness within the field of view of the microscope (circles). The thinnest region possesses a thicknesses of around 190 nm. With coatings applied, we can investigate the membrane in a home-build confocal microscope either in reflection (b) or in fluorescence (c), here with a broad detection window (600 nm to 800 nm). The dark fringes in reflection (b) occur at thicknesses of the diamond membrane, where the cavity is resonant to the green excitation light, yielding enhanced fluorescence (c) at the same positions. Calculating $\xi$ in dependence of the membrane thickness and the dipole emission wavelength (d), we can estimate the total emission spectrum of the cavity at exemplary positions (marked \textit{x} and \textit{o}). The measured fluorescence spectra at these positions (e) matches the expectations from the calculated $\xi$ very well, veryfing the underlying model.}
\label{fig:BareMembraneLM}
\end{figure*}

The optical constants chosen for the optimization in Sec. \ref{sec:optimization} are taken from \textit{McPeak et al.} \cite{McPeak2015}. The optical properties of the thick silver layer we actually fabricate is close to their values, we measure $n+\ii k = 0.07 + 4.10\ii$ at 620 nm. For the thin silver layer, however, we measure deviating optical properties, $n+\ii k = 0.15 + 3.95\ii$, resulting in a reduced performance of the antenna. The measured $n=1.45$ for silica at 620 nm is close to the literature value.
Based on these measured optical properties and coating thicknesses of the evaporated layers, we can calculate $\xi$ for the actually fabricated device for different membrane thicknesses $t_0$ and dipole emission wavelengths $\lambda$ with $d=27.5$ nm and $\vartheta=54.7^\circ$ fixed. This calculation is summarized in the plot in Figure \ref{fig:BareMembraneLM} d). 
Before investigating single emitters, we verify this plot and thus the model in general with measurements of the cavity resonances at different positions and thus thicknesses of the membrane.
\par
\subsubsection{Cavity resonances}
Exchanging the fluorescence filters in the detection path of the confocal microscope with a neutral density filter enables us to spatially map the reflectance and thus the cavity resonances for the excitation laser emitting at 516 nm.
Figure \ref{fig:BareMembraneLM} (b) shows a scan over the whole membrane with applied coatings. It can clearly be seen that the detected photon rate, in this case corresponding to the reflected laser power, drops at sharply defined positions on the membrane, forming fringes of low reflectance.
When we perform the same scan again with a 610 nm longpass filter instead of a neutral density filter, we observe the inverted situation in fluorescence, see Figure \ref{fig:BareMembraneLM} (c): Whenever the laser hits a resonance, the PL increases. The PL originates from background fluorescence of the diamond membrane that couples to the cavity modes. As we know that the cavity has to be resonant to the laser at these positions, we can directly deduce the membrane thickness at these positions from the plot in Figure \ref{fig:BareMembraneLM} (d). From the derived thickness, we can look up the other resonance wavelengths of the cavity for this thickness and compare it against measured spectra of the enhanced background fluorescence. As can be seen from Figure \ref{fig:BareMembraneLM} (e), the exemplary shown measured spectra match the theoretical prediction of the cavity resonances, confirming the theoretical model based on the measured properties of the layers.
\par
\begin{figure*}[h]
\centering
\includegraphics[width=\linewidth]{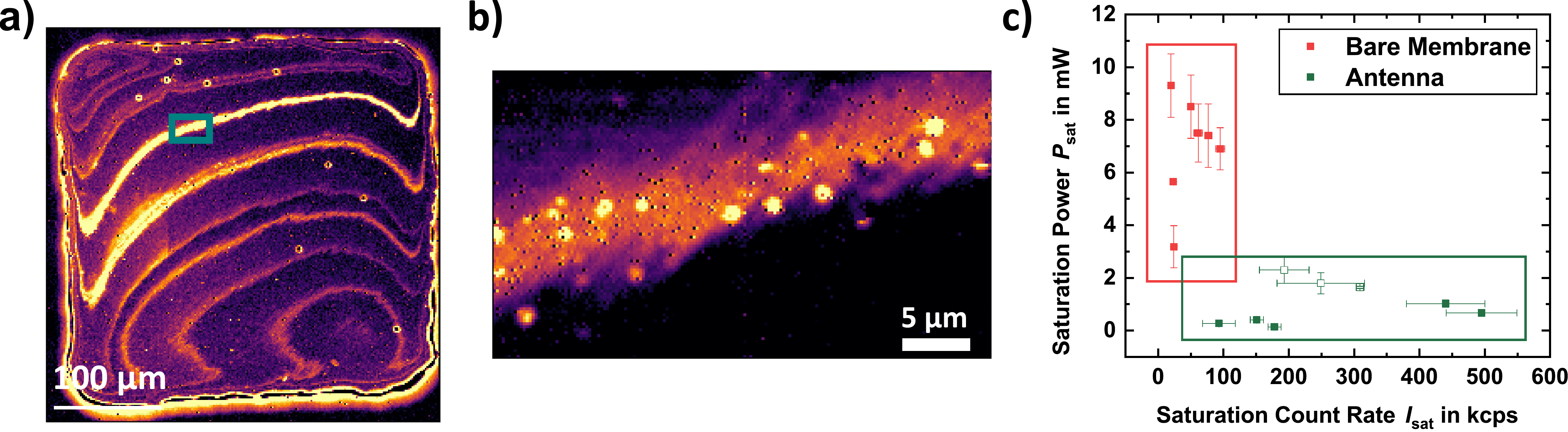}
\caption{(a) PL map with a filter bandwidth from 610 nm to 650 nm in detection. The bright line in the upper half of the membrane indicates the membrane thickness, at which the cavity is resonant to both the excitation laser and the detection bandwidth. This is the working point where the cavity is operating as an optical antenna under off-resonant excitation. (b) A finer scan at this working point (green frame in (a)) reveals single spots distributed around the cavity resonance. (c) Photodynamics of single emitters with a reasonable single photon purity close to the antenna working point. These emitters exhibit saturation powers well below and saturation rates well above the the bare membrane case. By accident, some emitters where measured with a narrower detection bandwidth (hollow squares, 615 nm - 625 nm). }
\label{fig:finalStatistics}
\end{figure*}
Setting the detection bandwidth to 610 - 650 nm, we measure a PL map as shown in Figure \ref{fig:finalStatistics} (a). In contrast to the map in Figure \ref{fig:BareMembraneLM} (c), we now see pairs of resonances, approaching each other spatially as the membrane thickness and thus the cavity length increases, until they finally merge in a resonance marking the optical antenna working point at a membrane thickness of around 609 nm. 
Each pair of resonances consists of one resonance for the excitation laser wavelength and one overlapping with the detection bandwidth.
In the former case, the visible PL originates from background fluorescence of the diamond membrane which feeds the cavity resonance. The cavity resonance, on the other hand, overlaps to certain degree with the narrow detection filter window, generating a detectable signal.
These are the resonances we also see in Figure \ref{fig:BareMembraneLM} (c).
In the latter case, the cavity resonances are centered within the narrow detection bandwidth and the weak coupling of the laser is enough to induce a significant PL.
\par
The brightest line in the PL map in \ref{fig:finalStatistics} (b) at around 609 nm diamond thickness indicates the spatial overlap of a cavity resonance for the excitation laser as well as for the detection bandwidth, yielding a strong PL signal. This is the working point of the optical antenna where we finally look for single emitters and their emission enhancement. From Figure \ref{fig:BareMembraneLM} (d), we can extract a maximum $\xi=0.214$ at $t_0 = 608.6$ nm. For the emitters we measured in the bare membrane at an approximate thickness of around 190 nm, we calculate a collection factor of $\xi = 0.022$. Thus, we expect a nearly 10-fold theoretical enhancement of $\xi$ due to the optical antenna, compared to the bare membrane.
\subsubsection{Enhancement of color center emission}
We search for single \SnVs at the working point, i.e. close to the bright fringe in Figure \ref{fig:finalStatistics} (a). A finer scan, as can be seen in Figure \ref{fig:finalStatistics} (b) reveals the existence of several individual spots which can be investigated for single photon emission.
When photon autocorrelation shows a dip below 0.5 and can be fully explained with uncorrelated background and detection time jitter, we measure the saturation behavior and add it to the statistics. Figure \ref{fig:BareMembraneMeas} d), e) and f) show an exemplary spectrum, saturation measurement and photon autocorrelation measurement for a single \SnV at the working point. From the spectrum in Figure \ref{fig:BareMembraneMeas}  d) it is inferable, that for this emitter the spectral overlap of the cavity resonance and ZPL is very good, yielding a high saturation photon rate $I_\text{sat}$. Additionally, the laser power $P_\text{sat}$ needed to saturate the emitter drops to very small values.
As mentioned above, we would expect an around 10-fold enhancement of the photon rate at the antenna working point. However, as the emitters are not always spatially located exactly at the proper membrane thickness but may be slightly offset, we expect to see a reduced enhancement on average. 
\par
Figure \ref{fig:finalStatistics} (c) finally summarizes the main results of this work. Shown in red are the measured saturation parameters for the emitters in the bare membrane, with photon rates well below 100 kcps and saturation powers of on average 6 mW.
Shown in green are all reliably identified single emitters in the antenna around the working point of $d=609$ nm, yielding single photon emission rates up to 500 kcps.
It is directly visible that $P_\text{sat}$ is tremendously reduced for all emitters found. To quantify the effect on $I_\text{sat}$, we have to look at the underlying numbers in more detail. The lowest (highest) $I_\text{sat}$ of an emitter in the bare membrane is 20 kcps (95 kcps), whereas it is 93 kcps (495 kcps) for an emitter in the antenna, yielding an enhancement of 4.7 (5.2). The measured average saturation count rate on all 7 emitters in the bare membrane (antenna) is $46\pm27$ kcps ($271 \pm 183$ kcps), yielding an average enhancement of $5.9$, matching the expectations well.
Due to the small number of emitters investigated, the error bars on this statistic are fairly large. 
The reason for this low number of emitters that were suitable to be taken into account is not only the fact that not all of the actually measured emitters were satisfying the restrictions on the $\gtwo$ value. Moreover, during the measurement of the antenna we found the coatings to severely degenerate. All of the emitters in the statistics of the optical antenna have been measured before a break in order to examine the first data. After this two months break, the silver layers showed severe color changes, indicating a degradation, although we applied a silica protection coating. As a consequence, no more reliable measurements could be carried out. Additionally, we tried to recycle the membrane by removing the silica as well as the silver coating via wet chemical processes, which unfortunately led to a destruction of the thin diamond membrane. 
\section{Discussion \& Outlook} 
Thin free-standing single crystal diamond membranes are a key requirement for the creation of many advanced nanophotonic structures to enhance single photon emission or spin-photon interaction for color centers. We demonstrated the successful fabrication of such a membrane starting with commercially-available high-purity diamond material. Creation of single color centers inside the membrane was shown to be straightforward using ion implantation techniques and subsequent annealing.
Transforming the thin diamond membrane into a cavity-based optical antenna requires in principle only established thin-film deposition and analysis technologies. Finally, the investigation of the color centers in the bare membrane and subsequently in the antenna has been carried out thoroughly, in both cases taking great care to unambiguously identify single emitters.
As a main result, we found a significant enhancement of single photon emission and reduction of saturation powers of the color centers coupled to the optical antenna compared to the uncoupled case in the bare diamond membrane, yielding count rates up to around half a million photons per second from a single SnV$^-$ center. The theoretical framework predicts these enhancements in good agreement to the measured data. Additionally, the occurrence of resonances in PL as well as in reflectance are well explained by the model. 

We consider this work as a proof of concept with many options for improvement: Although the emitters proved to be photostable even at high excitation powers, we observe randomly switching background fluorescence from the membrane, severely reducing the signal-to-background ratio of the actual emitter PL. This blinking background is currently the main limitation for the comparably low single photon purity. We observe it independently of the color center PL on all samples that have been plasma etched and subsequently annealed and cleaned as described in this work. We do further not observe it when focusing deep into the diamond, indicating that it originates from the surface. It has been shown that a hydrogen-termination \cite{Maier2000} or a disordered oxygen-termination \cite{Sangtawesin2019} of the diamond surface may act as a source of charge traps, leading to random charge fluctuations upon laser excitation. Promising options to remove this potential source of the blinking background fluorescence are more sophisticated post-processing techniques, e.g. oxygen annealing \cite{Sangtawesin2019} or a treatment with a purely inductively coupled oxygen plasma \cite{FavarodeOliveira2015,Radtke2019}. Both methods have been shown to introduce a highly-ordered oxygen termination, severely enhancing the charge state stability and spin coherence of shallow color centers.
Moreover, the vanishing linear background contribution in the measured saturation curves gives rise to a different emission dynamic compared to the SiV$^-$ center. For more sophisticated applications using SnV$^-$ centers, a thorough investigation of this potentially novel photophysic will be mandatory.

As already indicated in the main text, the coating of the diamond membranes appeared to be sufficient for our purposes in the beginning, yet the degradation of the silver layers proved us wrong. A possible explanation for this degradation is the weak chemical bonding between the silver and the silica layer: It is well known that the adhesion between noble metals and oxides such as silica is comparably weak already directly after deposition and degrades over time\cite{Gadkari2005}. Especially thin silver layers are known to be susceptible to dewetting even at room temperature\cite{McPeak2015}. A commonly used adhesion promotion can be established via a thin titanium or chromium layer between the oxide and the noble metal. In future work, we may thus need to include such an additional layer in our antenna design. As the comparably low reflectance of chromium or titanium may reduce the final performance\cite{Huang2021}, one could also think of protecting the upper silver layer with an additional gold layer. 
Lastly, the thickness gradient in the membrane provides us a varying cavity length, but limits the areas in which the cavity operates as an optical antenna. This raises the question which gradient can be tolerated at most. Our model covers only completely planar stacks and it is obvious that a wedged membrane will at some point severely change the properties of the cavity. Under the assumption that the gradient in first order only shifts the resonance wavelength, we can extract this shift from our model to be 14 nm for a change in thickness of the membrane of 10 nm. With the confocal microscope, we collect only the PL from a spot with a diameter of around 800 nm. If we accept a change in resonance wavelength of 6 nm within this spot diameter, equal to the linewidth of the SnV center, we find an upper bound for the gradient of 4.4 nm change of thickness per µm change in lateral position. With the estimated 1.8 nm per µm gradient we achieve here, we were thus able to show the successful antenna operation. Anyway, the long-term goal is surely to produce membranes with a well defined thickness and with well defined gradients on areas spanning hundreds of micrometers to fully harness the potential of this planar design. A first step towards dry etching processes that actively planarize diamond membranes has already been shown \cite{Heupel2019}, demonstrating the feasibility of such approaches.
\par
Besides these technical imperfections just summarized, we have successfully demonstrated a design for an optical antenna based on a planar Fabry-Pérot cavity which is able to boost the emission rate of a single \SnV center even in the worst scenario (case III in table \ref{tab:Optimization}) to around half a million photons per seconds at the detectors in saturation. In future work, we want to focus on near-resonant and resonant excitation (case II in table \ref{tab:Optimization}), which would allow for photon rates on the order of several million counts per second. Additionally, a reduction of the background fluorescence and thus an improvement of the single photon purity should be feasible by employing surface treatments as described above. Already at room temperature, such high photon rates together with a reasonable single photon purity would pave the way towards applications in quantum metrology or quantum sensing.
For the implementation of an efficient spin-photon-interface, however, cryogenic temperatures are unavoidable. Because of the monolithic design and thus the absence of moving parts, the design presented here may also perform well at low temperatures. Operation of the antenna under cryogenic conditions and resonant excitation consequently will fully harness the potential of this comparably simple nanophotonic design.
\section{Acknowledgements}
We thank B. Lägel and S. Wolff (Nano Structuring Center, NSC, University of Kaiserslautern) for helpful discussions on evaporation and use of their facilities. We further thank R. Hensel (Leibniz Insitute for New Materials, INM, Saarbrücken) for access to the \textit{Oxford} ICP RIE, and E. Neu, M. Challier, O. Opaluch and J. Görlitz for helpful discussions on etching and spectroscopy.
This research has been funded by the European Quantum Technology Flagship Horizon 2020 (No. H2020-EU1.2.3/2014-2020) under Grant No. 820394 (ASTERIQS); by the EMPIR programme co-financed by the Participating States and from the European Union’s Horizon 2020 research and innovation programme (project 17FUN06 SIQUST) and by the German Federal Ministry of Education and Research (Bundesministerium für Bildung und Forschung, BMBF) within the project Q.Link.X (Contract No. 16KIS0864).

\section{Data availability}
The data that support the findings of this study are available from the corresponding author upon reasonable request.
\clearpage
\bibliography{Antenna2019}

\end{document}